\documentstyle[12pt,psfig]{article}
\begin{document}


\def\mys#1{\Sigma_{ } { }_{\bf #1}}
\def\bc{\begin{center}}
\def\ec{\end{center}}
\def\vp{\vspace*{0.2cm}}
\def\vm{\vspace*{-0.2cm}}

\begin{center}

\vspace{1.0cm}

{\large \bf 
Critical Exponents of the Statistical Multifragmentation Model
}

\vspace*{0.5cm}

{\bf P.T. Reuter$^{1}$ and K.A. Bugaev$^{1,2}$ }

\vspace*{0.7cm}

$^1$ Institut f\"ur Theoretische Physik,
Universit\"at Frankfurt, Germany\\
$^2$ Bogolyubov Institute for Theoretical Physics,
Kyiv, Ukraine\\

\date{\today}

\end{center} 

\begin{abstract}
For the statistical  multifragmentation model the critical indices $\alpha^\prime, \beta,\gamma^\prime, \delta$ are calculated as functions of the Fisher parameter $\tau$. It is found that these indices have different values than in Fisher's droplet model. Some peculiarities of the scaling relations are discussed. The basic model predicts for the index $\tau$  a narrow range of values, $1.799< \tau < 1.846$, which is consistent with two experiments on nuclear multifragmentation.   

\end{abstract}

\vspace*{0.5cm}

\noindent
\hspace*{1.5cm}\begin{minipage}[t]{14.cm}
{\bf Key words:}
1st order liquid-gas phase transition,
 critical indices, scaling laws, nuclear matter
\end{minipage}

\vspace*{0.5cm}

\noindent
\hspace*{1.5cm}{\bf PACS} numbers: 25.70. Pq, 64.60. Fr, 21.65.+f, 64.70. Fx

\vspace*{0.5cm}

\begin{center}
{\large \bf I. Introduction} 
\end{center} 

\vm

The statistical multifragmentation model (SMM) \cite{Bo:95,Gro:97}
rather
successfully explains the multiple production 
of intermediate mass fragments \cite{Bot:95,Ag:96}. 
On the other hand it is also an example of 
a statistical model exhibiting a phase transition (PT)
of the liquid-gas type \cite{Ag:96,Po:95}.
Therefore,  the SMM   
was studied extensively in order 
to clarify not only 
the relationship  between  multifragmentation
and the nuclear liquid-gas PT,
but also to elucidate the  connection of the SMM
to both Fisher's droplet model (FDM) \cite{Fi:67} and 
the lattice gas model \cite{Gu:95a,Gu:95b}.  
For the last purpose a simplified version of the SMM
was suggested \cite{Gu:98,Gu:99}. 

In recent works \cite{Bug:00,Bug:01} an exact analytical solution of 
a simplified version of the SMM was found in the thermodynamic limit. 
This model is mathematically very similar to the original 
FDM. However, it differs from the FDM in two aspects:
 (i) the nonzero volume  of the nucleons
 is taken into account, and (ii) the free surface energy
is non-negative.
These two main differences generate distinct results
concerning the 
properties of the critical point: 
the SMM predicts the existence of a critical point 
for the parameter $\tau \le 1$ and a tricritical point
for $ 1 < \tau \le 2$ (see below), whereas for the
FDM this parameter
is assumed to be larger, $ \tau > 2$.
Such a difference is not surprising since the FDM
considers the droplets to have zero volume, i.e., 
its validity is restricted to low particle densities. 
Consequently, the FDM conclusions regarding the critical behavior must  
be considered as mainly suggestive in relation to the behavior of real 
fluid systems or of more realistic models.
Therefore, it is necessary to study how the critical  
indices of the SMM relate to those of the FDM.
It will be shown that the critical indices in the SMM and the FDM have 
different values, i.e., these models belong to 
different universality classes.

Another important issue is the comparison of the critical indices of the SMM
with
the corresponding values found in numerical studies 
\cite{El:00} and in experiments.
As we shall show the experimental data for the index $\tau$ 
are consistent with our predictions for the SMM. 

The paper is organized as follows.
In Sect.~II a brief description of the simplified version of the SMM is 
given. In Sect.~III some useful formulae are derived and the critical indices of this model 
are found. The scaling laws and   the relations between  the
 critical indices of the SMM  are discussed in Sect.~IV. 
The conclusions are given in Sect.~V.

\vp

\begin{center}
{\large \bf II. The Statistical Multifragmentation Model} 
\end{center} 

\vm

In Refs. \cite{Bug:00,Bug:01}  a simplified version of the SMM is solved analytically with the method suggested in \cite{Go:81}. The interaction of the fragments is taken into account by replacing the total volume $V$ by the free volume $V_f \equiv V- bA$, where $A$ is the number of nucleons in the system and $b$ is the eigenvolume of a nucleon. In the grand canonical ensemble the pressure of the system is found for the liquid and gaseous phase
\begin{equation}\label{pg}
p_l(T,\mu) = \frac{\mu + W(T)}{b}~, \hspace{0,2cm} p_g(T,\mu)=T{\cal F}(p_g,T,\mu)\,,
\end{equation}
respectively, where 
\begin{eqnarray}\label{Fs}  
{\cal F}(p_g,T,\mu) &\equiv&
g(T)
\left[z_1 e^{\frac{\nu-W(T)}{T}}+ \mys{0}(\varepsilon, \nu) \right]\\
{\rm and} \hspace{0.7cm} \mys{q}( \varepsilon, \nu) & \equiv &
\sum_{k=2}^{\infty}~k^{q-\tau}~e^{\textstyle \frac{\nu}{T} k - \frac{a(\varepsilon)}{T} k^\sigma}\,, 
\end{eqnarray}
with $g(T) \equiv \left[ mT /(2\pi)\right]^{3/2}$. The mass of a nucleon is denoted as $m$, $z_1 = 4$ is the number of spin-isospin states of a nucleon, $\nu \equiv \mu + W(T)- bp_g$ is the shifted chemical potential of the gaseous phase and $W(T)$ is the free energy per particle inside of a fragment.
The critical temperature of the system is denoted as $T_c$ and  $\varepsilon \equiv (T_c - T)/T_c$ defines the reduced temperature. The actual parameterization of $W(T)$ and the free surface tension $a(\varepsilon)$ can be found in Ref. \cite{Bug:00,Bug:01}. In this paper it is assumed that $W(T)$ and all its derivatives are regular functions of temperature and that the free surface tension obeys $a(\varepsilon)=a_{\rm o}  \varepsilon^\zeta$ for small positive values of $\varepsilon$ and  vanishes for $\varepsilon \leq 0$. The surface area of a fragment of $k$ nucleons is taken in the general form $k^\sigma$. In the SMM one usually chooses $\zeta = 5/4$ and $\sigma = 2/3$, but in the following treatment $\zeta \ge 1 $ and $0<\sigma<1$ will be regarded as free parameters. 

The particle  densities  in the liquid and in the gaseous phase  
are 
\mbox{
$
 \rho_l  \equiv  
\left(\partial  p_l/\partial \mu\right)_{T}
=1/b
$}
and
\mbox{
$\rho_g \equiv
\left(\partial  p_g/\partial \mu\right)_{T}=
\rho_{id}/( 1 + b \rho_{id} )
$,}
respectively. The density  
\begin{eqnarray}\label{rhoid}
\rho_{id}(T,\mu) \equiv g(T)
\left[z_1 e^{\frac{\nu - W(T)}{T}} 
 +  \mys{1}(\varepsilon, \nu )\right]
\end{eqnarray}
has the meaning of the ideal gas density in the limit $b \rightarrow 0$.

When both pressures coincide,  \mbox{ $p_l=p_g$}\,, $\nu = 0$\,, the liquid-gas PT occurs. This equality  defines the PT equilibrium curve $\mu^*(T)$ in the $(T,\mu)$-plane. For $T<T_c$ the surface tension $a(\varepsilon)$ is nonzero and the PT is of 1st order, since $\rho_{id}(T,\mu)<\infty$ and consequently $\rho_g<\rho_l$ on the curve $\mu^*(T)$.
The critical point at $T=T_c$ corresponds to the critical density $\rho_c = \rho_l = 1/b$. Since at the critical point $\nu=0$ and the surface tension vanishes, it follows from Eqs. (\ref{pg}) and (\ref{Fs}) that for $\tau \leq 1$ the function ${\cal F}$ is divergent and, therefore, $p_c$ is infinite. For $1 < \tau \le 2$ one finds similarly that $p_c < \infty$\,, $\rho_c=1/b$ and a 2nd order PT for $T>T_c$ (see Ref. \cite{Bug:00,Bug:01} for more details). For $\tau \ge 2$ on the phase equilibrium line both ${\cal F}(p^*(T),T,\mu^*(T))$ and $\rho_{id}(T,\mu^*(T))$ are finite for any finite temperature, i.e., the mixed phase region extends formally to infinite temperatures. Therefore, this case exhibits no critical point at all in contrast to the FDM and is not considered in the following.

\vp

\begin{center}
{\large \bf III. The Critical Indices of the SMM}
\end{center}

\vm

The critical exponents $\alpha^\prime, \, \beta$ and $\gamma^\prime$ describe the temperature dependence of the system near the critical point on the coexistence curve $\nu = 0 $
\begin{eqnarray}\label{alpha}
c_\rho &~ \sim ~& 
\left\{
\begin{tabular}{ll}
$\mid \varepsilon \mid^{-\alpha}$\,, \hspace*{0.95cm} ${\rm for} ~ \varepsilon  < 0 $~, \\
$\varepsilon^{-\alpha^\prime}$\, ,\hspace*{1.25cm} ${\rm for} ~ \varepsilon  \ge 0$~,
\end{tabular}
\right.
\\ \label{beta}
\Delta\rho &~ \sim ~ & 
\varepsilon^{\beta}\,,   \hspace*{2.cm}	 {\rm for} ~ \varepsilon  \ge 0 ~,
\\ \label{gamma}
\kappa_T &~ \sim ~&  
\varepsilon^{-\gamma^\prime}\,,\hspace*{1.7cm} 	 {\rm for} ~  \varepsilon  \ge 0 ~,
\end{eqnarray}
where $\Delta\rho \equiv \rho_l - \rho_g$ defines the order parameter, $c_\rho \equiv \frac{T}{\rho} \left(\frac{\partial s}{\partial T} \right)_\rho$ denotes the specific heat at fixed particle density and $\kappa_T \equiv \frac{1}{\rho} \left(\frac{\partial \rho}{\partial p} \right)_T$ is the  isothermal compressibility.
The shape of the critical isotherm for $\rho \leq \rho_c$ is given by the critical index $\delta$
\begin{equation}\label{delta}
p_c - \tilde{p}  ~\sim ~
(\rho_c  - \tilde{\rho})^{\delta}		\hspace*{1.1cm} {\rm for} ~  \varepsilon = 0~. 
\end{equation}
The tilde indicates $\varepsilon = 0$ hereafter.

The calculation of $\alpha$ and $\alpha^\prime$ requires the specific heat $c_{\rho}$. With the formula \cite{Ya:64}
\begin{equation}\label{crho}
\frac{c_{\rho}(T, \mu)}{T}~= ~ \frac{1}{\rho}\left(\frac{\partial^2 p}{\partial T^2} \right)_{\rho} - \left(\frac{\partial^2 \mu}{\partial T^2} \right)_{\rho}
\end{equation}
one obtains the specific heat on the PT curve by replacing the partial derivatives by the total ones \cite{Fi:70}. The latter can be done for every state inside or on the boundary of the mixed phase region.
For the chemical potential $\mu^*(T) = bp^*(T) - W(T)$ one gets
\begin{equation}\label{crho*}
\frac{c_{\rho}^*(T)}{T}~ = ~ \left( \frac{1}{\rho}- b \right) \frac{ {\rm d}^2 p^*(T)}{ {\rm d} T^2} + \frac{{\rm d}^2 W(T)}{{\rm d} T^2}~,
\end{equation}
where the asterisk indicates the condensation line ($\nu = 0$) hereafter. 
Fixing $\rho = \rho_c = \rho_l = 1/b$ one finds $c_{\rho_l}^*(T) = T\frac{ {\rm d^2}W(T)} {{\rm d}T^2}$ and, hence, obtains similarly  for both $\alpha$ and $\alpha^\prime$
\begin{equation}\label{resalpha}
\alpha ~= ~\alpha^\prime ~= ~ 0~.
\end{equation}
To calculate $\beta$ and $\gamma^\prime$ the behavior  of the series
\begin{equation}\label{I}
\mys{q}( \varepsilon, 0)  ~ = ~ 
\sum_{k=2}^{\infty}~k^{q-\tau}~e^{ - A\varepsilon^\zeta k^\sigma }
\end{equation}
for small positive values of $\varepsilon$ should be analyzed \mbox{$(A \equiv a_{\rm o}/T_c)$}.
In the limit $\varepsilon \rightarrow 0$ the function $\mys{q}( \varepsilon, 0)$ remains finite, if $\tau > q+1$, and diverges otherwise. For $\tau = q+1$ this divergence is logarithmic. The case $\tau < q + 1$ is analyzed in the following.

With the substitution ${\mbox z_{k}\equiv k\left( A \varepsilon^\zeta \right)^{1/\sigma} }$ it follows
\begin{equation}\label{series}
\mys{q}( \varepsilon, 0)~= ~\left(A  \varepsilon^\zeta \right)^{\textstyle \frac{\tau - q }{\sigma}}~ 
\sum_{k=2}^{\infty}~
z_k^{q-\tau}~e^{\textstyle -z_k^\sigma}~.
\end{equation}
Since \mbox{$\Delta z_{k}= \left( A \varepsilon^\zeta \right)^{1/\sigma} \equiv \Delta z$}, 
the series on the r.\,h.\,s. of (\ref{series}) converges to an integral for $\varepsilon \rightarrow 0$
\begin{equation}
\mys{q}( \varepsilon, 0)~= ~\left(A  \varepsilon^\zeta \right)^{\textstyle \frac{\tau-q-1}{\sigma}}
\hspace*{-3 mm}\int \limits_{2 \left( A  \varepsilon^\zeta \right)^{\frac{ 1}{\sigma}}}^{\infty}
\hspace*{-3 mm} {\rm d}z~
z^{q-\tau}~e^{\textstyle -z^\sigma} ~.
\end{equation}
The assumption $q-\tau > -1$ is sufficient to guarantee the convergence of the integral at its lower limit.
Because of the finite value of the above integral
one obtains the following result for $\varepsilon \rightarrow 0$
\begin{equation}\label{Prop1}
\mys{q}( \varepsilon, 0) \sim \left\{ 
\begin{array}{ll}
\varepsilon^{\textstyle\, \frac{\zeta}{ \sigma }(\tau - q- 1) }\,,
&
{\rm if} ~ \tau < q + 1 \,,  \\
\ln\mid\varepsilon\mid\,,  
&
{\rm if} ~ \tau = q + 1 \,,
\\
\varepsilon^{\,0}\,, 
&
{\rm if} ~  \tau > q + 1 \, .
\end{array}
\right.
\end{equation}
For the calculation of $\delta$ one obtains similarly
\begin{equation}\label{Prop2}
\mys{q}( 0, \tilde{\nu})~\sim ~\left\{ 
\begin{array}{ll}
\tilde{\nu}^{\textstyle\, \tau - q- 1}\,,
&
{\rm if} ~ \tau < q +1\,,  
\\
\ln\mid\tilde{\nu}\mid\,,  
&
{\rm if} ~ \tau = q + 1 \,,
\\
\tilde{\nu}^{\,0}\,,
&
{\rm if} ~  \tau > q + 1 \,.
\end{array}
\right.
\end{equation}

From  
$\Delta\rho=\frac{1}{b}\frac{1}{1+b\rho_{id}}$ it follows $\Delta\rho \sim \rho_{id}^{*{~-1}}$ for $\varepsilon \rightarrow 0$\,. 
Therefore, inserting $q=1$ into Eq. (\ref{Prop1}) immediately gives  $\beta$
\vspace{-0.3cm}
\begin{eqnarray}\label{resbeta}
\beta
~=~\frac{\zeta}{\sigma}\left(2 - \tau \right)~.
\end{eqnarray}

From $\rho_g = \rho_{id}/( 1 + b \rho_{id} )$ and Eq. (\ref{rhoid}) it follows that
\begin{equation}\label{kappaepsilon}
\kappa_T~=~\frac{g(T)}{T} \frac{\rho_g}{\rho_{id}^3}
 \left[ z_1 e^\frac{\nu - W(T)}{T} 
+ \mys{2}(\varepsilon, \nu)\right]~. 
\end{equation}
Using Eq. (\ref{Prop1}) one easily finds $\gamma^\prime$ from Eq. (\ref{kappaepsilon})
\begin{equation}\label{resgamma}
\gamma^\prime ~= ~\frac{2\zeta}{\sigma}\left( \tau - \frac{3}{2}  \right) ~.
\end{equation}
Near the critical point it follows $\rho_c - \tilde{\rho} \sim 1/ \tilde{\rho}_{id}$. With (\ref{rhoid}) 
one finds using Eq. (\ref{Prop2})~
$\rho_c - \tilde{\rho}  \sim   \left[\mys{1}(0, \tilde{\nu})\right]^{-1}  \sim  \tilde{\nu}^{2 - \tau}$\,.
Substituting $1=e^{\frac{-\tilde{\nu}}{2T_c}  k}e^{\frac{\tilde{\nu}}{2T_c}  k}$ yields for small values of $\tilde{\nu}$
\begin{eqnarray}\label{peps}
p_c - \tilde{p} &~\cong~& g(T_c) T_c \left[ -\frac{z_1}{T_c} \tilde{\nu}
+ \sum_{k=2}^{\infty} 
k^{ - \tau} \left(1- e^{ \frac{\tilde{\nu}}{T_c}  k} \right)\right]  \nonumber\\
&~\cong~& g(T_c) \left[ -z_1 \tilde{\nu}
- \tilde{\nu} \sum_{k=2}^{\infty} 
k^{1 - \tau}~ e^{ \frac{\tilde{\nu}}{2T_c}  k} \right] \nonumber \\
& \sim & \tilde{\nu}~ \mys{1}\left(0, { \textstyle \frac{\tilde{\nu}}{2}} \right) ~ \sim ~  \tilde{\nu}^{\tau - 1}~, \hspace{1cm} {\rm for}~  \tilde{\nu} < 0 ~.
\end{eqnarray}
Combining the result above with the expression found for $\rho_c - \tilde{\rho}$ one obtains the critical index $\delta$
\begin{equation}\label{resdelta}
\delta ~ = ~ \frac{\tau - 1}{2 - \tau}~,
\end{equation}
which is independent of the choice of $\zeta\,,~\sigma$ and $\tau$.

\vp

\bc
{\large \bf IV. Scaling Relations of the SMM}
\ec

\vm

In the special case $\zeta = 2\sigma$
 the well-known exponent inequalities proven for real gases by
\begin{eqnarray}
\label{F1}
{\rm Fisher \cite{Fi:64}:}& \hspace{.5cm}			\alpha^\prime + 2\beta + \gamma^\prime ~& \ge~2 ~,\\
\label{G}
{\rm Griffiths \cite{Gri:65}:}& 		\alpha^\prime + \beta(1 + \delta) & \ge~ 2 ~,\\
\label{L}
{\rm Liberman \cite{Li:66}:}&			\gamma^\prime + \beta(1-\delta) & \ge ~0 ~,
\end{eqnarray}
are fulfilled exactly for any $\tau$. (The corresponding exponent inequalities for magnetic systems are often called Rushbrooke's, Griffiths' and Widom's inequalities, respectively.) For $\zeta > 2\sigma $,  Fisher's and Griffiths' exponent inequalities are fulfilled as inequalities and for $\zeta < 2\sigma $ they are not fulfilled. The contradiction to Fisher's and Griffiths' exponent inequalities in this last case is not surprising. This is due to the fact that in the present version of the SMM the critical isochore $\rho = \rho_c = \rho_l$ lies on the boundary of the mixed phase to the liquid. Therefore, in  expression (2.13) in Ref. \cite{Fi:64} for the specific heat only the liquid phase contributes and the proof of Ref. \cite{Fi:64} following (2.13) cannot be applied for the SMM. Thus, the exponent inequalities (\ref{F1}) and (\ref{G}) have to be modified for the SMM. Using Eqs. (\ref{resalpha}), (\ref{resbeta}), (\ref{resgamma}) and (\ref{resdelta})  one finds the following scaling relations
\begin{eqnarray}
\alpha^\prime + 2\beta + \gamma^\prime  =   \frac{\zeta}{\sigma} 
\hspace{0.5cm} {\rm and} \hspace{0.5cm}
\alpha^\prime + \beta(1 + \delta)  =  \frac{\zeta}{\sigma} ~.
\end{eqnarray}
Liberman's exponent inequality (\ref{L}) is fulfilled exactly for any choice of $\zeta$ and $\sigma$.

Since the coexistence curve of the SMM is not symmetric with respect to $\rho = \rho_c$, it is interesting with regard to the specific heat to consider the difference $\Delta c_\rho(T) \equiv c^*_{\rho_g}(T) - c^*_{\rho_l}(T)$, following the suggestion of Ref. \cite{Fi:70}. With Eq. (\ref{crho*}) and noting that $1/\rho_g^* - b = 1/\rho_{id}^*$ it follows
\begin{eqnarray}\label{Drho}
\Delta c_\rho(T) ~=~ \frac{T}{\rho_{id}^*(T)} \frac{ {\rm d}^2 p^*(T)}{ {\rm d} T^2}~.
\end{eqnarray}
The most divergent term in Eq. (\ref{Drho}) yields for $\zeta>1$
\begin{equation}\label{alphas}
\alpha^\prime_s ~= ~
\left\{ 
\begin{array}{ll}
\vspace{0.1cm}2 - \frac{\zeta}{\sigma}\,,
& 
{\rm if} ~ \tau < \sigma + 1 \,,  \\
2 - \frac{\zeta}{\sigma}( \sigma + 2 - \tau)\,,	
&
{\rm if} ~  \tau \ge \sigma + 1 \,.
\end{array}
\right.
\end{equation}
Then it is $\alpha_s^\prime > 0$ for $\zeta/\sigma <2$.
Thus, approaching the critical point along any isochore within the mixed phase region except for $\rho = \rho_c = 1/b$ the specific heat diverges for $\zeta/\sigma <2$ as defined by $\alpha^\prime_s$ and remains finite for the isochore $\rho = \rho_c = 1/b$. This demonstrates the exceptional character of the critical isochore in this model.

In the special case that $\zeta = 1$ one finds $\alpha^\prime_s = 2 - 1/\sigma$ for $\tau \leq 1 + 2 \sigma$ and $\alpha^\prime_s = -\beta$ for $\tau > 1 + 2 \sigma$.
Therefore, using $\alpha_s^\prime$ instead of $\alpha^\prime$, the exponent inequalities (\ref{F1}) and (\ref{G}) are fulfilled
exactly if $\zeta >1$ and  $\tau \leq \sigma + 1$ or if $\zeta =1$ and  $\tau \leq 2\sigma + 1$. In all other cases (\ref{F1}) and (\ref{G}) are fulfilled as inequalities.
Moreover, it can be shown that the SMM belongs to the universality class of real gases for $\zeta >1$ and $\tau \ge \sigma + 1$.

The comparison of 
the above derived formulae for the critical exponents of the SMM
for $\zeta = 1$
with  
those obtained within the FDM 
(Eqs. 51-56 in \cite{Fi:67}) shows that these models
belong to  different  universality classes (except for the singular case $\tau = 2$). 
 
Furthermore, one has to note that for $\zeta = 1\,,~\sigma \leq 1/2$ and $1 + \sigma < \tau \leq 1 + 2 \sigma$ the critical exponents of the SMM coincide with those of the exactly solved one-dimensional FDM with non-zero droplet-volumes \cite{Fi:70}. 

For the usual parameterization of the SMM \cite{Bo:95} 
one obtains with $\zeta = 5/4$ and $\sigma = 2/3$ the exponents
\begin{eqnarray}
\alpha^\prime_s ~&=& ~\left\{ 
\begin{array}{ll}
\vspace{0.2cm} \frac{1}{8}\,, 
&
{\rm if} ~ \tau < \frac{5}{3}   \\
\frac{15}{8}\tau - 3\,, 
&
{\rm if} ~  \tau \ge \frac{5}{3}  
\end{array}
\right.~,\hspace*{0.5cm}
\beta~=~
\frac{15}{8}\left(2 - \tau \right)~, \nonumber\\
\gamma^\prime ~&=& ~
\frac{15}{4}\left( \tau - \frac{3}{2} \right)~,\hspace*{2.cm}
\delta ~ = ~ \frac{\tau - 1}{2 - \tau}~. 
\end{eqnarray}
The critical indices of the nuclear liquid-gas PT were determined 
from the multifragmentation of gold nuclei \cite{Eos:94} and
found to be close to those ones of real gases.  
The method used to extract the critical exponents 
$\beta$ and $\gamma^\prime$  in Ref. \cite{Eos:94} was, however,
found to have large uncertainties of about 25 per cents \cite{Bau:94}.
Nevertheless, those results allow us  
to estimate the value of $\tau$
from the experimental values of the critical exponents  of real
gases taken with large error bars.
Using Eqs. (\ref{resbeta}), (\ref{resgamma}) and (\ref{resdelta}) one can generalize the exponent relations of Ref. \cite{Fi:70}
\begin{eqnarray}
\label{ntau1}
\tau ~= ~ 2 - \frac{\beta}{ \gamma^\prime + 2 \beta} ~~~ {\rm and} \hspace{0.5cm} 
\tau ~= ~ 2 - \frac{1}{1 + \delta}
\end{eqnarray}
for arbitrary $\sigma$ and $\zeta$. 
Then, one obtains with \cite{Hu} $\beta = 0.32-0.39$\,, $\gamma^\prime = 1.3-1.4$ and $\delta = 4-5$ the estimate $\tau =  1.799 - 1.846$. 
This demonstrates also that the value of $\tau$ is rather 
insensitive to the special choice of $\beta\,,~\gamma^\prime$ and $\delta$.
Theoretical values for $\beta\,,~\gamma^\prime$ and $\delta$ for Ising-like systems within the renormalized $\phi^4$ theory \cite{Zi:98} lead to the narrow range $\tau = 1.828 \pm 0.001$\,.

The critical indices of the SMM were studied numerically in Ref.  
\cite{El:00}. The version V2 of Ref. \cite{El:00} corresponds  
precisely to 
our model with $\tau = 0$, $\zeta = 5/4$ and $\sigma = 2/3$,
but their results contradict to our analysis.
Their results for version V3 of Ref. \cite{El:00} are in contradiction with
our proof presented in Ref. \cite{Bug:00}. There it was shown that for non-vanishing surface energy (as in version V3)
the critical point does not exist at all.
The latter was found in \cite{El:00}
for the finite system
and the critical indices were analyzed.
Such a strange result
is, probably, due to
finite volume effects, although
some doubts about the validity of
the methods used to extract the critical indices (especially, $\tau$)
remain.

It is widely believed that the effective value of $\tau$ defined 
as $\tau_{\rm eff} \equiv  - \partial \ln n_k(\varepsilon) / \partial \ln k$ 
 attains
its minimum at the critical point (see references in \cite{EOS:00}).
This has been shown for the version of the FDM with the constraint of sufficiently small surface tension $a \cong 0$ for $T\ge T_c$ \cite{Pan:83} and also can be seen easily for the SMM. Taking the SMM fragment distribution  
$n_k(\varepsilon) = g(T) k^{-\tau} \exp[{\textstyle \frac{\nu}{T} k 
- \frac{a(\varepsilon)}{T} k^\sigma }] \sim k^{- \textstyle \tau_{\rm eff}}  $ 
one finds
\begin{equation}\label{teff}
\tau_{\rm eff}~ = ~ \tau - \frac{\nu}{T} k + \frac{\sigma a(\varepsilon)}{T} k^\sigma  
\quad \Rightarrow \quad \tau = \min( \tau_{\rm eff}) ~,
\end{equation}
where the last step follows from the fact that  
the inequalities 
$a(\varepsilon) \ge 0$\,, $\nu \leq 0$ become equalities  
at the critical point $\nu = a(0) = 0$. 
Therefore,
the SMM predicts
that the minimal value of $\tau_{\rm eff}$ corresponds to the critical point.

In the E900 $\pi^-+$\,Au multifragmentation experiment \cite{Be:99} 
the ISiS collaboration measured
the dependence of
$\tau_{\rm eff}$ upon the excitation energy and 
found the minimum value ${\rm min}(\tau_{\rm eff} ) \cong 1.9$ 
(Fig.\,5 of Ref. \cite{Be:99} ). 
Also the EOS collaboration \cite{EOS:00} performed an analysis of the minimum of $\tau_{\rm eff}$ on Au\,+\,C multifragmentation data. The fitted $\tau_{\rm eff}$, plotted in Fig.\,11.b of Ref. \cite{EOS:00} versus the fragment multiplicity, exhibits a minimum in the range 
${\rm min}(\tau_{\rm eff}) \cong 1.8 - 1.9$\,. 
Both results contradict the original FDM \cite{Fi:67}, but agree 
with the above estimate of $\tau$ for real gases and for Ising-like systems in general.

\vp

\bc
{\large \bf V. Conclusions}
\ec

\vm

The critical indices of an exactly soluble version of the SMM are derived.   
The inclusion of excluded volume effects generates 
a principal difference between the SMM and the FDM -- these models
belong to different universality classes. 
It is found that
for the critical exponent $\alpha^\prime$ obtained 
traditionally on the critical isochore
Fisher's and Griffiths' exponent inequalities are broken. 
The modification of the definition of $\alpha^\prime$
in the spirit of Ref. \cite{Fi:70} recovers both 
scaling relations.  
The scaling laws (\ref{ntau1}) connecting $\tau$ with $\beta$, $\gamma^\prime$ 
and $\delta$ are generalized for arbitrary $\sigma$ and $\zeta$. 
The ranges $\tau = 1.799 - 1.846$ for real gases and $\tau = 1.828 \pm 0.001$ for Ising-like systems are estimated. 
It is shown that experimental data on nuclear multifragmentation 
agree well with these ranges. 
Therefore, it is possible that the values for the exponent $\tau < 2$ 
seen in the data evidence that the 
critical point of the nuclear liquid-gas phase transition   
has been reached experimentally.

\vspace*{0.5cm}

{\bf  Acknowledgments.}  
The authors 
 thank M.E.~Fisher, M.I.~Gorenstein, I.N.~Mishustin,
J.~Polonyi and D.H. Rischke
 for useful discussions and valuable comments. 
K.A.B. is thankful to  the Alexander von Humboldt Foundation
for financial support. 
The research described in this publication was made possible in part by
Award No. UP1-2119 of the U.S. Civilian Research \& Development
Foundation for the Independent States of the Former Soviet Union
(CRDF).


\vspace{-0.3cm}

\begin{thebibliography}{99}


\bibitem{Bo:95}
J.P. Bondorf {\it et al.}, 
Phys. Rep. {\bf 257} (1995) 131;

\bibitem{Gro:97}
D.H.E. Gross, Phys. Rep. {\bf 279} (1997) 119.

\bibitem{Bot:95}
A.S. Botvina {\it et al.}, Nucl. Phys. {\bf A 584} (1995) 737;

\bibitem{Ag:96}
M. D'Agostino et. al., Phys. Lett. {\bf B 371} (1996) 175;
Nucl. Phys. {\bf A 650} (1999) 329.

\bibitem{Po:95}
J. Pochodzalla et. al., Phys. Rev. Lett. {\bf 75} (1995) 1040.

\bibitem{Fi:67}
M.E. Fisher, Physics {\bf 3} (1967) 255.

\bibitem{Gu:95a}
J. Pan and S. Das Gupta, Phys. Lett {\bf B 344} (1995) 29.

\bibitem{Gu:95b}
J. Pan and S. Das Gupta, Phys. Rev. {\bf C 51} (1995) 1384. 

\bibitem{Gu:98}
S. Das Gupta and A.Z. Mekjian, Phys. Rev. {\bf C 57}
(1998) 1361.

\bibitem{Gu:99}
S. Das Gupta, A. Majumder, S. Pratt and A. Mekjian,
{\it Preprint} {\bf nucl-th/9903007} (1999).

\bibitem{Bug:00}
K.A. Bugaev, M.I. Gorenstein, I.N. Mishustin and W. Greiner,
Phys. Rev. {\bf C62} (2000) 044320; 
{\it Preprint} {\bf nucl-th/0007062} (2000).

\bibitem{Bug:01}
K.A. Bugaev, M.I. Gorenstein, I.N. Mishustin and W. Greiner,
Phys. Lett. {\bf B 498} (2001) 144; 
{\it Preprint} {\bf nucl-th/0103075} (2001).

\bibitem{El:00}
J. B. Elliott and A. S. Hirsch,
Phys. Rev. {\bf C61} (2000) 054605.

\bibitem{Go:81}
M.I. Gorenstein, V.K. Petrov and G.M. Zinovjev, Phys. Lett.
{\bf B 106} (1981) 327.

\bibitem{Ya:64}
C.N. Yang and C.P. Yang, Phys. Rev. Lett. {\bf 13} (1964) 303

\bibitem{Fi:70}
M.E. Fisher and B.U. Felderhof, Ann. of Phys. {\bf 58} (1970) 217

\bibitem{Fi:64}
M.E. Fisher, J. Math. Phys. {\bf 5} (1964) 944

\bibitem{Gri:65}
R.B. Griffiths, J. Chem. Phys. {\bf 43} (1965) 1958

\bibitem{Li:66}
D.A. Liberman, J. Chem. Phys. {\bf 44} (1966) 419

\bibitem{Eos:94}
M. L. Gilkes {\it et al.},
Phys. Rev. Lett. {\bf 73} (1994) 1590  

\bibitem{Bau:94}
W. Bauer and W.A. Friedman,
{\it Preprint} {\bf  nucl-th/9411012} (1994) 

\bibitem{Hu}
K. Huang, {\it Statistical Mechanics}, J. Wiley, New York (1987)

\bibitem{Zi:98}
R. Guida and J. Zinn-Justin, J. Phys. Math. Gen. {\bf 31} (1998) 8103-8121


\bibitem{EOS:00}
J. B. Elliott {\it et al.} (the EOS Collaboration),
Phys. Rev. {\bf C62} (2000) 064603.

\bibitem{Pan:83}
A. D. Panagiotou {\it et al.}, Phys. Rev. Lett. {\bf 52} (1983) 496

\bibitem{Be:99}
L. Beaulieu {\it et al.}, Phys. Lett. {\bf B463} (1999) 159




\end{thebibliography}
\end{document}